\documentclass[sigplan,screen]{acmart}

\usepackage{stfloats}
\usepackage{enumitem}
\usepackage{booktabs}
\usepackage{pgf-pie}

\usepackage{multicol}
\usepackage{listings}
\usepackage{color}
\usepackage{xcolor}

\definecolor{dkgreen}{rgb}{0,0.6,0}
\definecolor{gray}{rgb}{0.5,0.5,0.5}
\definecolor{mauve}{rgb}{0.58,0,0.82}
\AtBeginDocument{%
  \providecommand\BibTeX{{%
    \normalfont B\kern-0.5em{\scshape i\kern-0.25em b}\kern-0.8em\TeX}}}

\setcopyright{acmlicensed}
\acmConference[SecTL '23]{Secure and Trustworthy Deep Learning Systems }{July 10--14, 2023}{Melbourne, VIC, Australia}
\renewcommand\footnotetextcopyrightpermission[1]{}
\setcopyright{acmcopyright}
\copyrightyear{2023}
\acmYear{2023}
\acmDOI{10.1145/3591197.3591308}

\acmBooktitle{Secure and Trustworthy Deep Learning Systems (SecTL '23), July 10--14, 2023, Melbourne, VIC, Australia}
\acmPrice{15.00}
\acmISBN{979-8-4007-0181-8/23/07}

\begin{document}

\title{Beyond the Model: Data Pre-processing Attack to Deep Learning Models in Android Apps}


\author{Ye Sang}
\email{ysan0006@student.monash.edu}
\affiliation{%
  \institution{Monash University}
  \city{Melbourne}
  \state{Victoria}
  \country{Australia}
  \postcode{3168}
}

\author{Yujin Huang}
\email{Yujin.Huang@monash.edu}
\affiliation{
  \institution{Monash University}
  \city{Melbourne}
  \state{Victoria}
  \country{Australia}
  \postcode{3168}
}

\author{Shuo Huang}
\email{Shuo.huang1@monash.edu}
\affiliation{
 \institution{Northwestern Polytechnical University}
 \city{Xi'an}
 \country{China}
 \postcode{710129}
  \institution{, Monash University}
  \city{Melbourne}
  \state{Victoria}
  \country{Australia}
  \postcode{3168}
}

\author{Helei Cui }
\email{chl@nwpu.edu.cn}
\affiliation{
 \institution{Northwestern Polytechnical University}
 \city{Xi'an}
 \country{China}
 \postcode{710129}
}
\authornote{Corresponding author}


\begin{abstract}
The increasing popularity of deep learning (DL) models and the advantages of computing, including low latency and bandwidth savings on smartphones, have led to the emergence of intelligent mobile applications, also known as DL apps, in recent years.
However, this technological development has also given rise to several security concerns, including adversarial examples, model stealing, and data poisoning issues.
Existing works on attacks and countermeasures for on-device DL models have primarily focused on the models themselves. However, scant attention has been paid to the impact of data processing disturbance on the model inference. 
This knowledge disparity highlights the need for additional research to fully comprehend and address security issues related to data processing for on-device models.
In this paper, we introduce a data processing-based attacks against real-world DL apps. In particular, our attack could influence the performance and latency of the model without affecting the operation of a DL app. 
To demonstrate the effectiveness of our attack, we carry out an empirical study on 517 real-world DL apps collected from Google Play.
Among 320 apps utilizing MLkit, we find that 81.56\% of them can be successfully attacked. 

%  216 commented by Alex, the number should be 320 instread of 260
The results emphasize the importance of DL app developers being aware of and taking actions to secure on-device models from the perspective of data processing.

\end{abstract}

\maketitle
\section{Introduction}
\label{sec:intro}
Deep learning has demonstrated significant effectiveness in various domains, such as face detection \cite{24, 25, 26}, natural language understanding \cite{huang2023training,huang2021robustness}, and speech recognition \cite{zhang2018deep}. On the other hand, mobile apps have brought a new level of convenience to people's daily lives, enabling activities such as reading, chatting, and banking on-the-go. To enhance the functionality of mobile apps, many developers have incorporated deep learning techniques to add artificial intelligence features such as image classification, face recognition, and voice assistants. These deep learning models can be hosted in the cloud or locally on the device within the app.

% Based on the known comparable performance of deep learning (DL) models, its have been applied in various fields such as face recognition \cite{24, 25, 26}, object detection \cite{23}, and image labeling \cite{27, 28}. As a result, many App developers have started integrating such models into their Apps to provide users with more convenience, what we call DL app. In this case, the latency and bandwidth issues encountered in data transferring from client to server could be addressed compared with traditional apps with AI service.

The main difference between a DL app with on-device DL models and one with on-cloud DL models is that the former performs the inference task locally while the latter needs to upload the data to the remote server. 
Considering the fact that uploading data requires the network’s stability and the server’s bandwidth, the on-device models gradually attract attention from the research community and industry. Several advantages will be brought by incorporating DL models into a mobile app. Firstly, privacy issues can be addressed as sensitive data do not need to transfer to the server, especially regarding medical and health tasks. Second, the model inference will be finished on the client rather than on the server, which save data transferring time and bandwidth and thus make the inference process efficient. Finally, the on-device model could perform customized tasks according to the behavior of the users via the local re-training, which improves user experience and also preserve user privacy\cite{1}. 

\begin{figure}[t]
  \centering
   \includegraphics[width=0.8\linewidth]{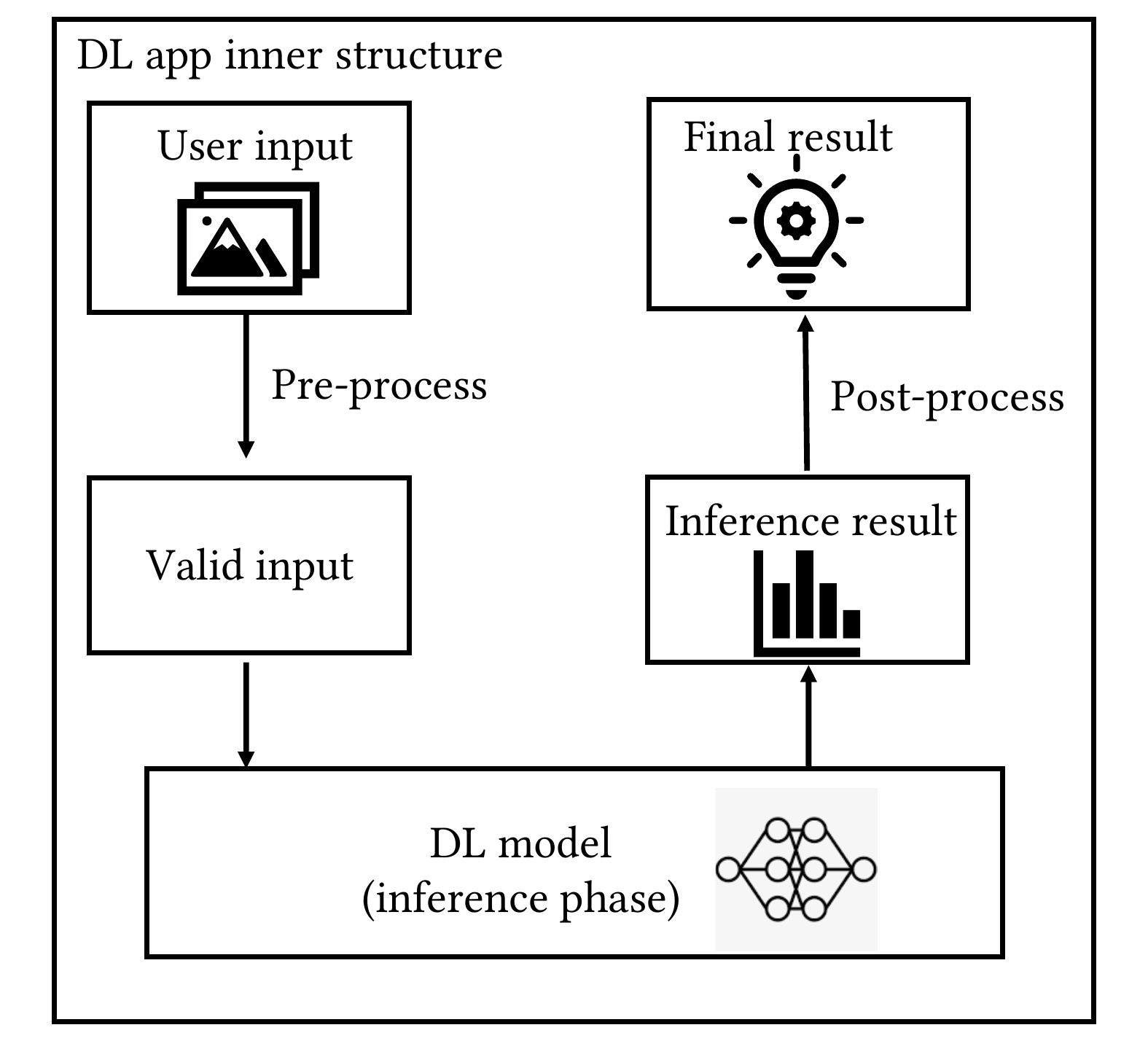}
   \caption{A typical inner structure in the DL apps.}
   \label{fig:1}
   \vspace{-0.2cm}
\end{figure}

However, unlike the more robust defense against adversarial attacks on the server, the on-device models are be more vulnerable to attacks as such models are inevitably stored in user devices. There are several works~\cite{1,2,3,4,7,21,22,huang2022smart,wang2023beyond} have explored the robustness of the on-device model. Among these works, most of them focus on the model itself. Expressly, they aim to reduce the performance of the on-device model by either modifying model structure or retraining a substitute model, the prerequisite of such attacks require strong knowledge of DL, no matter the black-box or white-box attack. Moreover, when the model is extremely complex, it may be difficult to execute such adversarial attacks.

In contrast to previous work, we propose a novel attack against DL apps through disrupting the data processing procedure.
Figure \ref{fig:1} shows the common structure of a DL app.
As seen, the essential step of a DL app is to pre-process data (e.g., resize, normalization, standardization) before performing the inference task. Hence, instead of targeting at the on-device model, we can manipulate the data pre-processing of a DL app to achieve attack.
To do so, we first summarize the general pre-processing rules of DL apps from a large number of DL apps.
Subsequently, we identify a DL app using a set of pre-defined model naming suffix.
Once we have knowledge of the general pre-processing rules and DL app identification, we can inject the malicious codes into DL apps by using reverse engineering to interrupt the pre-processing process, which could influence the result of the model inference.
After injecting the malicious codes into the pre-processing procedure of a DL app, we recomiple it to produce a compromised DL app that can be downloaded by any potential user. 
Note that we only focus on the TensorFlow Lite models in this work because they occupy most of DL apps~\cite{7}. Plus, according to the empirical study \cite{12}, 70.3\% DL apps take image as inputs when implementing the AI service, while 21.3\% take audio and text as inputs. Therefore, in the paper we only take the DL apps involved in image into account. Meanwhile, our attack can also be extended to natural language processing (NLP) tasks due to the similar pre-processing between text and vision tasks. For example, data normalization such as word embedding is an essential manner during the pre-processing in NLP tasks, and previous study \cite{52} has also elucidated the impact of varying data normalization techniques in natural language processing (NLP) tasks. Consequently, it is possible to perturb such preprocessing methods in order to influence the performance of NLP models.
%% commented by alex

There are two significant obstacles to executing this attack against real-world DL apps. The code obfuscation makes it difficult to find the desired code because the first and most important step in our methodology is to analyse the code and summarise the general rules using static analysis. The objective of obfuscation is to obscure the class, method, and variable names as much as possible while preserving the inner logic of the protocols.
An example of code obfuscation shown in Figure  \ref{obfus}. However, the obfuscation only targets the variable's name and not its value. In addition, initialising the DL model in a DL app requires loading the model locally via the model file path, which cannot be encrypted. By locating the code used to load the model, we could acquire the context of the pre-processing, post-processing, and model invocation method.

Another difficulty is permitting the modified App service to behave normally but make errors when conducting DL-related services. In other words, the core of our attack consists of designing and inserting malevolent codes in the correct location. To address this challenge, we locate the critical class and method used for data processing based on the data control flow, in accordance with the general data processing principles outlined at the outset. Ultimately, we design and insert malicious code into the target method.

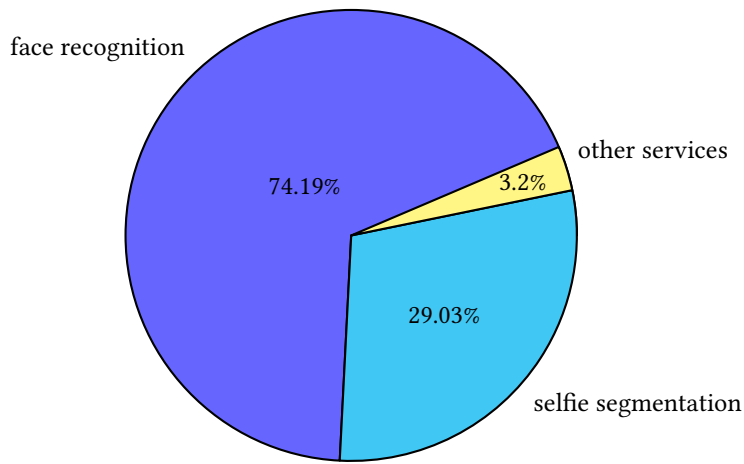
\begin{figure}[t]
\centering
\begin{tikzpicture}
    dc tag/.style={align=center,text width=5cm},
  \pie{74.19/face recognition, 29.03/selfie segmentation, 3.2/other services}
\end{tikzpicture}
\caption{The AI service distribution in our collected DL apps. Note that there may be overlapped between the services.}
\label{app_dis}
\end{figure}

We crawled 517 DL apps from Google Play and evaluate our attack on them in terms of latency and accuracy. Among these apps, our attack successfully injects the malicious code into 81.56\% (261/320) DL apps utilizing MLkit and achieve satisfactory performance. Specifically, we randomly selected the DL apps that were subjected to the attack and conducted the testing. As a result, it could found that a majority of these DL apps cannot perform normally regardless of the input provided. The AI service distribution for those DL apps is shown in Figure \ref{app_dis}

Our work presents this new attack to increase awareness within both the research community and industry regarding the need to protect on-device models. The contributions of this paper are as follows:
%% commented by ye
\begin{itemize}[leftmargin=*]
    \item We design and realize a new black-box data processing attack. This attack is simpler and more effective than previous works as it does not require attackers to comprehend the structure of the DL model in a DL app.

    \item We conduct an evaluation on its performance on 320 real-world DL apps with 81.56\% of them successfully attacked, demonstrating the effectiveness of our attack in terms of accuracy and latency. 
    We also make our source code\footnote{https://github.com/Alexender-Ye/attack-focused-on-pre-processing} available to the public to facilitate future research in this area.
\end{itemize}

The remainder of the paper is organized as follows. We describe
the preliminary knowledge in Section 2. We present our attack methodology in Section 3. We present
the experiment results of attacking real-world DL apps Section 4. We discuss the limitations and possible future work in Section 5. 

% 1) We design and implement a new black-box attack focused on data processing. This attack does not require the attackers to know about the structure of the DL model in an Android app and, thus, is more effective than previous works. 

% 2) This attack aims to insert malicious code into the original App and further influence the performance of DL models and cannot be influenced by obfuscation which is one of the most valuable mechanisms to protect the App. 

% 3) In this paper, we applied this attack to a real-world App acquired from apkpure~\cite{37} by using a crawler, demonstrating its feasibility. Moreover, we also test it in terms of accuracy and latency.

% \begin{lstlisting}[ language=Java,label = obfus,caption={As is depicted in the above listing, the class, function and variable names have been modified into simple character due to the obfuscation}]
% public final class c
% {
%     public c(androidx.profileinstaller.d.c c1, int i, b obj)
%     {
%         a = c1;
%         b = i;
%         c = obj;
%     }
% }

% \end{lstlisting}

\begin{figure}[ht]
\begin{multicols}{2}
\begin{lstlisting}[ language=Java, caption={The codes before the obfuscation}]
private void demoCode() {
    int[] temp = new int[100];
    int sum = 0;
    }

\end{lstlisting}
\columnbreak
\begin{lstlisting}[language=Java, caption={The codes after the obfuscation}]
private void a() {
    final int[] n2 = new int[100];
    int n4 = 0;
}
\end{lstlisting}
\end{multicols}
\caption{The performance of obfuscation}
\label{obfus}
\end{figure}

\begin{figure*}[htbp]
  \centering
%   \fbox{\rule{0pt}{2in} \rule{0.9\linewidth}{0pt}}
   \includegraphics[width=0.75\linewidth]{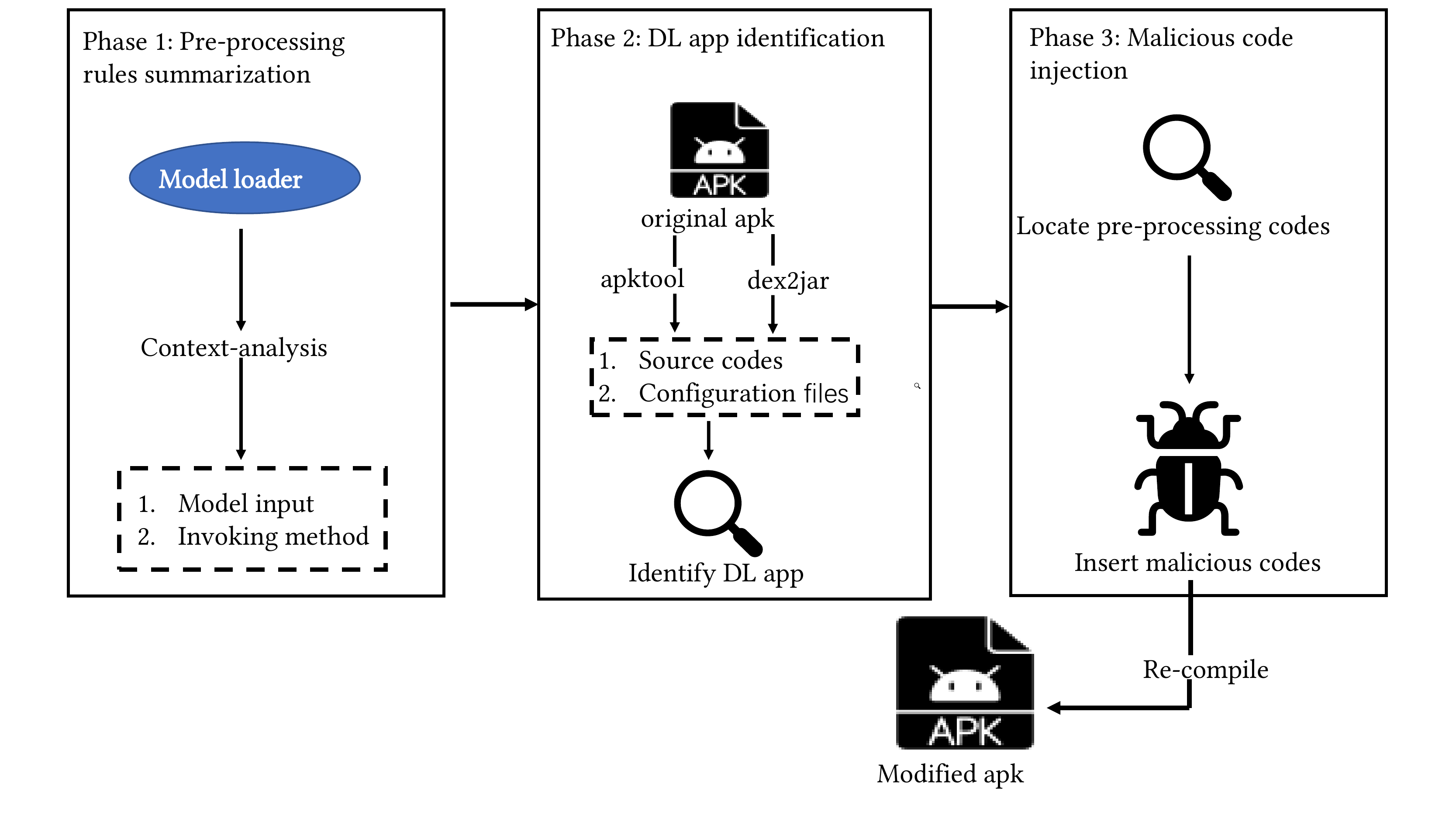}
   \caption{The overall workflow of our attack.}
   \label{fig:3}
   \vspace{-0.2cm}
\end{figure*}

\section{Preliminary}
\label{sec:related work}
\subsection{Mobile DL Frameworks}
With the development of computational power, smartphones gradually penetrate the public's daily life. In addition, the promotion of DL also contributes to making people live and work conveniently. Moreover, in recent years, the public's attention to data privacy has driven industries to achieve the inference task on the client end. Considering these, several on-device DL models have been rolled out by various vendors. In contrast to the traditional manner of performing the inference task on the server end, the on-device model is a pre-trained model that could finish the inference task on edge devices. There are some advantages brought through this way. First, there is no need for edge devices to send data to the server, indicating the saving of bandwidth, low latency, and privacy protection. Second, the on-device model could perform more customized compared to a general server model.

Therefore, several DL frameworks could allow the developers to deploy DL models onto the mobile phone, such as the TensorFlow Lite (TFLite) \cite{29}, PyTorch Mobile \cite{30}, Apache MXNet from ASF \cite{31}, Embedded Learning Library from Microsoft \cite{32}, Core ML 3 from Apple \cite{33}, and ncnn from Tencent \cite{34}. The availability of these frameworks saves the App vendors lots of effort. According to previous works \cite{4}, TFLite \cite{29} occupied 40\% of the number of App with DL models in 2018. Moreover, the related research \cite{5} also illustrated the performance of TFLite in model size, speed, and accuracy. Therefore, it is likely for the developers to use such kind of DL framework to deploy the model on edge devices.

\subsection{The Potential Threat for DL on the Mobile}

In combination with those mentioned above, it is also of great significance to focus on the security of DL in edge devices. Although the on-device model offers plenty of advantages, it also contributes to some new security issues. First, since the model is deployed on edge devices, the attackers are more likely to attack or even steal these models directly than before. For example, \cite{7} also demonstrated the similarity of 61.18\% of 62822 DL apps models are more than 80\% to the pre-trained model in TensorFlow Hub \cite{8}. Due to the high vulnerability of the pre-trained model, it is easy for adversaries to achieve a back-door attack \cite{9}. As a result, it might be easy for adversaries to attack these applications. Meanwhile, this research also mentioned that only 59\% of the DL apps take measures to protect their models \cite{6}. 

However, most attacks and the corresponding protections assume that the adversaries have a strong background in DL. As mentioned in this research \cite{10}, four system attack surfaces exist. Most of the current work focuses on the model perspective. On the contrary, we design a new black-box attack from a data processing perspective and apply it in real Apps.

\subsection{Obfuscation}
Due to the existence of reverse engineering, the developers use obfuscation to avoid leakage. Obfuscation is claimed by \cite{16} to disrupt reverse engineering. Although the apktool \cite{35}, one of the reverse-engineering approaches, could generate code up to 97\% similar to the source code according to this research \cite{11}, the obfuscation still makes it difficult for adversaries to understand the code. The principle of obfuscation is to obfuscate the class, method and variable name, leading to poor code readability. Notably, some developers are prone to convert the crucial part of source code to byte code. \cite{51} claims that obfuscation is one of the most effective methods to protect code, especially for those popular applications. Meanwhile, previous work \cite{50} also demonstrated the obfuscation was applied in approximate 50\% of the most popular apps with more than 10 million downloads. For example, as is shown in Figure \ref{obfus}, the function and variable names have been changed due to the obfuscation.

%% commented by alex

\section{Threat Model}
\paragraph{Attacker’s Goal.}

Given an Android application with DL model, the adversaries aim to make this application perform normally, but produce unexpected result while involving in specific AI service, such as face detection and selfie segmentation. Specifically, the adversaries will only inject the malicious codes into the modules involved in the data processing used for model inference.

\paragraph{Attacker’s Background Knowledge.}
This attack aim to explore if the disturbance of the data processing will affect the model performance. In contrast to previous work \cite{18,19,20,22} focused on the DL algorithm, the adversaries dont' know about the inner structure of the embedded DL model and have the access to the original training dataset. 

\paragraph{Attack Scenarios.}

Suppose there is an Android application based on DL. It is assumed that the adversaries can get access to the this application from market and repackage the modified application after injecting the malicious codes, which is one of the most commonly used method by adversaries \cite{45}. The modified application will be made available in some third-part market, because most of the third-part market seldom check the validity of and consistency of the application\cite{46,47,48,49}. In this case, when users download and install the application from these sources, it could provide the the adversaries with practicality to achieve the attack.

%% focus on cv

\section{Attack Methodology}
\label{sec:methodology}
The attack workflow as is shown in Figure \ref{fig:3} encompasses three steps: DL apps pre-processing rule summarization, DL apps identification, and DL apps malicious code injection. 
We first summarize the general pre-processing rules of DL apps from large-scale DL apps collected by previous works~\cite{7,12,38,39,40}.
Then, we introduce a mechanism to identify a (victim) DL app.
Finally, the summarized pre-processing rules are employed to attack the victim app.

\begin{figure}[hbp]
  \centering
%   \fbox{\rule{0pt}{2in} \rule{0.9\linewidth}{0pt}}
   \includegraphics[width=1\linewidth]{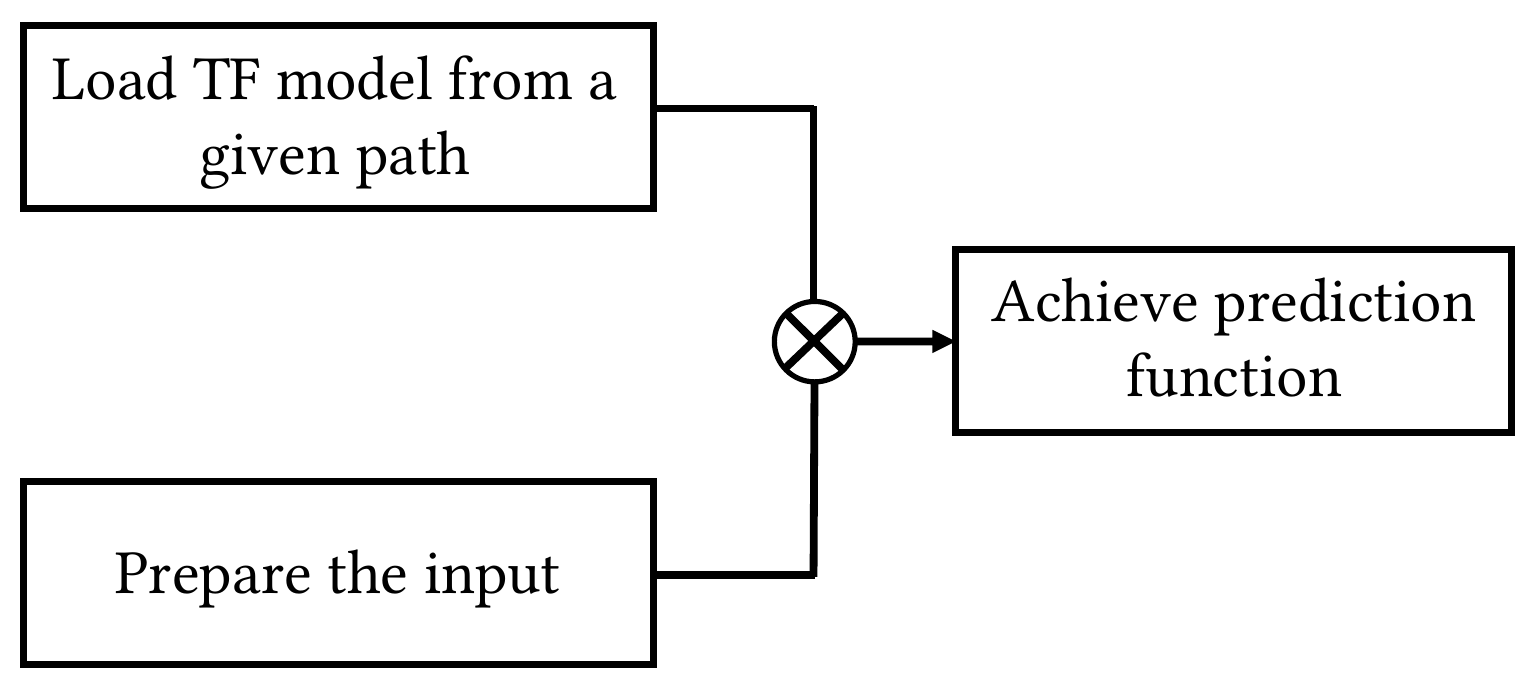}
   \caption{An example of the standard operation process of a DL app that adopts the TFLite model. Form this pipeline, it could be observed before calling inference function, the model object is expected to first load DL model file. Similarly, data object also need to be preprocessed.}
   \label{fig:4}
   \vspace{-0.3cm}
\end{figure}

\paragraph{DL App Pre-processing Rule Summarization.}
Data pre-processing is essential to model inference as raw inputs from mobile cameras need to be adjusted before utilisation.
Thus, once an attacker is aware of the pre-processing procedure of a DL app, the app is easily compromised.
Since manually observing the pre-processing rules of large-scale DL apps is expensive and time-consuming, we randomly select a small portion of DL apps across various categories from previous works~\cite{7,12,38,39,40} to conduct preliminary observations.
Consequently, we find that a DL app adopting TFLite models typically follows a standard operation process, an illustrative example is shown in Figure~\ref{fig:4}.
As seen, the on-device model is initially loaded using the TFLite Interpreter that accepts a model file path as the input. 
After model initialization, a model object capable of invoking the inference method is ready to use. 
According to the parameter of the inference method (predicte), the inferenceInput refers to the model input after pre-processing (e.g., rotation adjustment), and the pre-processing codes can be accessed through data flow analysis~\cite{41,42,43,44}.

On the basis of this observation, we design and implement an automated pipeline to summarize the general pre-processing rules from large-scale DL apps.
In particular, we define two criteria for locating the pre-processing code in our pipeline. First, a Computer Vision-related model typically accepts an image as input. Even if the input is a video, the developers will extract each frame as an image and pre-process it before feeding it to the on-device model, as the model cannot take streaming data as input. Note that the video-to-image conversion is not considered a preprocessing phase. 
Second, the Android API functions "createBitmap()", "createScaledBitMap", and "decodeResource ()" are frequently used to create the image object.
Based on these two criteria, we can trace the parameter used in the inference method until we locate the method used to generate a bitmap and consider these algorithms as pre-processing.
Nevertheless, many programmers prefer to use obfuscation to safeguard the most vital portion of their code. In this case, the essential section of code in a DL app is converted to bytecode, which cannot be directly located. To deal with this problem, we employ the technique proposed in \cite{14} to extract the internal logic through deobfuscation.
% The summarized pre-processing rules of DL apps are shown in Table~\ref{table2}.

\begin{figure}[t]
  \centering
%   \fbox{\rule{0pt}{2in} \rule{0.9\linewidth}{0pt}}
   \includegraphics[width=1\linewidth]{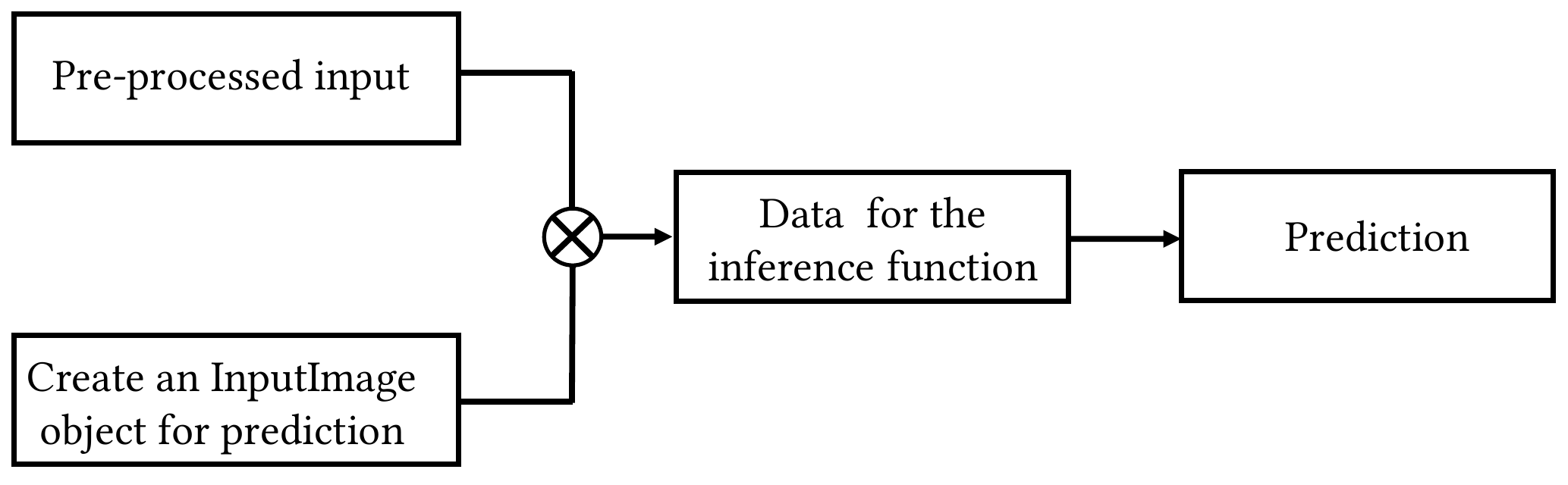}
   \caption{An example of the data pre-processing method employed by a DL app that adopts the TFLite model. After the pre-processing pharse, a new object of \textit{InputImage} used for inference function will be created.}
   \label{fig:6}
\vspace{-0.5cm}
\end{figure}

\paragraph{DL App Identification.}
As our attack focuses on DL apps adopting TFLite models, we employ reverse engineering followed by model recognition to identify such apps.
Given an Android APK, we use Apktool~\cite{35} to disassemble
it into nearly original form, including resource and configuration files.
After obtaining the disassembled APK, we determine whether it is a DL app via model recognition.

Expressly, we recognize a TFLite model based on a set of pre-defined model naming schemes, as shown in Table~\ref{table1}.
If the disassembled APK contains a file with any TFLite-related suffix, it is determined as a DL app.
However, some DL apps may encrypt their models (e.g., file name) due to security concerns.
Hence, we further improve our model recognition approach by adding TFLite library checking, which scrutinizes whether the source codes of the disassembled APK contain the TFLite API.
If so, the APK is identified as a DL app as well.
\begin{table}[htbp]
\begin{center}
\begin{tabular}{lc}
\toprule
DL FrameWork & Model File Suffix \\
\midrule
 & .tflite\\
 TensorFlow Lite& .tfl \\
 & .lite \\
\bottomrule
\end{tabular}
\end{center}
\caption{The naming schemes of TFLite models.}
\label{table1}
\vspace{-0.6cm}
\end{table}
\paragraph{DL App Malicious Code Injection.}
\label{malicious}
After knowing the pre-processing rules of DL apps and having the method for identifying DL apps, we now can inject malicious code against the pre-processing procedure of a DL app in order to compromise it. 
To this end, we first locate the pre-processing code of a DL app and then inject the malicious code into it to interrupt the pre-
processing procedure.
Finally, we recompile the modified app code back to its original form to produce a compromised DL app.
Recall that a DL app adopting TFLite models normally follows a standard pre-process procedure.
Therefore, we can locate the pre-processing code based on the pre-process procedure.
For the sake of consistent presentation, we resue the illustrative example in Figure~\ref{fig:4} and dive into its pre-processing method, as shown in Figure \ref{fig:6}.
Observe that an object of \textit{InputImage} will be created and regarded as the parameter in the later reference method.
However, regardless of the methods employed, the inner logic in these static methods is to create an object of \textit{InputImage} according to the parameters it received, such as height, width, rotation degree and image format. 
In other words, if the parameter values associated with data processing are modified during the creation of the object, the data processing approach in the DL app will be invalid and further influence the latency and accuracy of inference. In this case, through the analysis of the creation of this object, it has been found that the total three constructors in \textit{InputImage} will configure the image format field to a different value across different constructors. 
For instance, in the bitmap constructor, the image format field will be -1, whereas in the byte buffer constructor, it will be 842094169 or 17, which can be determined using a Python script.

This procedure, however, is fraught with two difficulties. First, some constructor-related codes have been obfuscated, indicating that it will be difficult to locate them. The second is that the internal logic of the data processing code may be varied in different DL apps.
In other words, we are faced with the difficulty of proposing a general rule to alter the data processing code.

To address the first challenge, the obfuscation only obfuscates the class, method, and variable names from the third package, while leaving the values and the Android vanilla method name unchanged. Additionally, the obfuscated variable name of the same version of TFLite API is identical, indicating that it would likely match directly if we know the variable from existing files after obfuscation. Regarding the second challenge, however, various versions of TFLite API may have variable names and data processing logic that vary. For instance, in the most recent version, matrix manipulation will be implemented via the class textitAndroid.graphics.Matrix, while in some old versions are not. Moreover, the variable name is not the same between different versions. In this instance, semantic search allows us to trace the code, and we define the following three search strategies:

\begin{itemize}[leftmargin=*]
    \item Strategy 1: When a constructor contains an assignment of 842094169 or 17 to a field of the current class accompanied by a specific string, this one will be regarded as a constructor in \textit{InputImage}.

    \item Strategy 2: When a constructor contains the Android vanilla method \textit{getHeight()} and \textit{getWidth()} invoked by a parameter and is involved in an assignment of -1 to a field of the current class, this one will be regarded as a constructor in InputImage.

    \item Strategy 3: When a constructor contains an assignment of 35 to a field in the current class and is involved in the operation of the matrix, this one will be regarded as a constructor in \textit{InputImage}.
\end{itemize}

\label{oimage}
\begin{lstlisting}[caption={The image constructor method of the InputImage class. Note that line 13 is responsible for the rotation value of an \textit{InputImage} object, which means if we can locate and mofify it, the performance of the DL model will be influenced.}]
.method private constructor <init>(Landroid/media/Image;IIILandroid/g)
    .locals 1
    .line 4
    invoke-direct {pe}, Ljava/lang/object;-><init>()V
    const/16 p4,xb4
    invoke-static (p1],Lcom/google/android/gms/common/internal/Prec
    new-instance ve,Lcom/google/mlkit/vision/common/zzb...
    invoke-direct {v0,p1},Lcom/google/mlkit/vision/common/zzb...
    iput-object v0,pe,Lcom/google/mlkit/vision/common/InputImage...
    iput p2,pe,Lcom/google/mlkit/vision/common/InputImage;->zzd:I
    iput p3,p0,Lcom/google/mlkit/vision/common/InputImage;->zze:I
    iput p4,p0,Lcom/google/mlkit/vision/common/InputImage;->zzf:I
    const/16 p1, x23
    iput p1,p0,Lcom/google/mlkit/vision/common/InputImage;->zzg:I
    iput-object p5,p0Lcom/google/mlkit/vision/common/InputImage;...

    return-void
.end method

\end{lstlisting}

% \begin{figure}[htbp]
%          \centering
%          \includegraphics[width=0.4\textwidth,height = 5cm]{con1.png}
%          \caption{Part of the code in InputImage class in terms of creating instance through image object.}
%          \label{fig:7.1}
%          \vspace{-0.2cm}
% \end{figure}

\label{obitmap}
\begin{lstlisting}[caption={The bitmap constructor method of the InputImage class. The codes in line 10 and 15 are responsible for creating a new object of InputImage which will be used in later inference function.},float = t]
.method private constructor <init>(Landroid/graphics/Bitmap;I)V
    .locals 1
    .line 1
    invoke-direct {pe], Ljava/lang/object;-><init>()V
    invoke-static (p1],Lcom/google/android/gms/common...
    move-result-object v0
    check-cast v0, Landroid/graphics/Bitmap;
    iput-object v0,p0, Lcom/google/mlkit/vision/common/InputImage;...
    .line 2
    invoke-virtual {p1}, Landroid/graphics/Bitmap;->getwidth()I
    move-result v0
    iput v0,p0, Lcom/google/mlkit/vision/common/InputImage:->zzd:I
    .line 3
    const/16 p2,6xb4
    invoke-virtual {p1}, Landroid/graphics/Bitmap;->getHeight()I
    move-result p1
    iput p1,pe, Lcom/google/mlkit/vision/common/InputImage;->zze:I
    iput p2,p0,Lcom/google/mlkit/vision/common/InputImage;->zzf:I
    const/4 p1, -0x1
    iput p1, pe, Lcom/google/mlkit/vision/common/Inputimage;->zzg:I
    const/4 p1, 0x0
    iput-object p1,p,0 Lcom/google/mlkit/vision/common/InputImage...
    return-void
.end method
\end{lstlisting}
% \begin{figure}[htbp]
%          \centering
%          \includegraphics[width=0.4\textwidth,height = 7cm]{con2.png}
%          \caption{Part of the code in InputImage class in terms of creating instance through bitmap.}
%          \label{fig:7.2}
%          \vspace{-0.2cm}
% \end{figure}

Once the pre-processin code is located, we now are able to perform malicious code injection.
In particular, we concentrate on modifying an image object’s height, width and rotation values.
By analyzing the \textit{InputImage} constructor, as shown in Listing~\ref{oimage}, the values of height, width and rotation will be assigned to the field of \textit{InputImage} object during its creation.
 Thus, if the value of these three parameters has been modified before the assignment operation, the generated object of InputImage will be changed and further influence the inference task. 
 Also, for the constructor associated with bitmap, as shown in the bottom part of Listing \ref{obitmap}, we could replace the getHeight() and getWidth(), which cannot be obfuscated with other values and modify the second parameter representing the rotation value. Once we have injected the malicious code, we will recompile the modified file to produce a compromised DL app.

\section{Evaluation}
To evaluate the performance of our attack on real-world DL apps, we realize our attack based on the methods introduced in Section~\ref{sec:methodology} and conduct an empirical study to evaluate the proposed attack against a collection of Android applications collected from the official Google Play market.

% \begin{figure}[t]
%          \centering
%          \includegraphics[width=0.4\textwidth,height=6.5cm]{pre1.png}
%          \caption{Part of resize codes in pre-processing phrase}
%          \label{fig:5.1}
% \end{figure}

% \begin{figure}[t]
%          \centering
%          \includegraphics[width=0.4\textwidth,height=10.5cm]{pre2.png}
%          \caption{Part of rotation adjustment codes in pre-processing phrase}
%          \vspace{-0.3cm}
%          \label{fig:5.2}
% \end{figure}

\begin{table}[htbp]
\begin{center}
\begin{tabular}{lc}
\toprule
Pre-processing & Examples\\
\midrule
Resize  & 480*360\\ 
Rotation adjustment  & 0,90,180,270\\ 
Normalize RGB values  & divide by 255\\ 
% No data processing  & \\
% Unidentified  & \\
\bottomrule
\end{tabular}
\end{center}
\caption{The summarized pre-processing rules of DL appps.}
\label{table2}
\vspace{-0.6cm}
\end{table}

\paragraph{Summarizing the General Pre-processing Rules of DL Apps.}
For the summarization of the pre-processing rules of DL apps, we employ the automated pipeline mentioned in Section~\ref{sec:methodology} to analyze 394 DL apps collected from previous works~\cite{7,12,38,39,40}.

Note that we do not analyze all DL apps in previous works as some of them cannot be properly executed by our pipeline.Table \ref{table2} shows the result of the pre-processing approaches. It could be observed that there are three common pre-processing ways. For example, the code in Listing \ref{resize_eg} aims to adjust the input resolution. If the difference between actual and desired values (640 * 320) is larger than a pre-defined value, the original input will be resized to a specific resolution. In the Listing \ref{rota_eg}, only the 0, 90, 180 and 270 will be regarded as proper rotation. Often, 0 will be set to the default value of rotation. In addition, the pre-processing of the RGB value normalization is typically achieved by dividing it by 255.
\label{rota_eg}
\begin{lstlisting}[ language=Java, caption={Calculates the correct rotation for the given camera id and sets the rotation in the parameters}]
WindowManager windowManager = (WindowManager) activity.getSystemService(Context.WINDOW_SERVICE);
int degrees = 0;
int rotation = windowManager.getDefaultDisplay().getRotation();
switch (rotation) {
    case 0:
        degrees = 0;
        break;
    case 1:
        degrees = 90;
        break;
    case 2:
        degrees = 180;
        break;
    case 3:
        degrees = 270;
        break;
}
CameraInfo cameraInfo = new CameraInfo();
Camera.getCameraInfo(cameraId, cameraInfo);
this.rotationDegrees = (cameraInfo.orientation + degrees) % 360;
int displayAngle = (360 - this.rotationDegrees) % 360; 
camera.setDisplayOrientation(displayAngle);
parameters.setRotation(this.rotationDegrees);
\end{lstlisting}
Interestingly, we find that most DL apps use a framework called firebase MLKit \cite{15} which could be directly used to achieve AI tasks, such as object detection, face detection and pose detection. 
In particular, the DL apps using MLKit will write the specific dependency of MLKit into the \textit{AndroidManifest.xml} file that could be obtained through apktool \cite{35}. For example, when a DL app uses the face detection of MLKit, the AndroidManifest.xml will contain \textit{com.google.mlkit.vision.fac\\e.internal.FaceRegistrar}. So, through this perspective, we can extract this information from .xml file for analysis and find that among the DL apps with MLKit, 90.32\% use the face detection,  5.76\% use the selfie segmentation and 3.22\% use the barcode identification services.
Moreover, we find that MLKit has developed a set of pre-processing rules for use with inference assignments. Not only does it require developers to resize the original input to enhance inference latency, but it also suggests adjusting the rotation degree to a particular value.

\label{resize_eg}
\begin{lstlisting}[language=Java, caption={An example code of resize. It will select the most suitable picture size, given the desired width and height. The method for selecting the best size is to minimize the sum of the differences between  the desired values and the actual values for width and height. },float = t]
List<SizePair> validPreviewSizes = generateValidPreviewSizeList(camera);  
SizePair selectedPair = null;
int minDiff = Integer.MAX_VALUE;
for (SizePair sizePair : validPreviewSizes) {
    Size size = sizePair.preview;
    int width = size.getWidth();
    int height = size.getHeight();
    int diff =
          Math.abs(width - 640) + Math.abs(height - 320);
    if (diff < minDiff) {
        selectedPair = sizePair;
        minDiff = diff;
    }
}
return selectedPair;
\end{lstlisting}

\paragraph{Collecting and Identifying DL Apps.}
For the collection of Android apps, we crawl 1,046 mobile apps from Google Play.
These apps involve a range of image-related categories (e.g., Medicine, Automobiles, and Finance).
We then identify the DL apps adopting TFLite models through the reverse engineering and model recognition mentioned in Section~\ref{sec:methodology}.
As a result, we obtain 517 TFLite-based DL apps that can be used as victim apps to evaluate our attack.

\begin{figure}[htbp]
  \centering
%   \fbox{\rule{0pt}{2in} \rule{0.9\linewidth}{0pt}}
   \includegraphics[width=\linewidth]{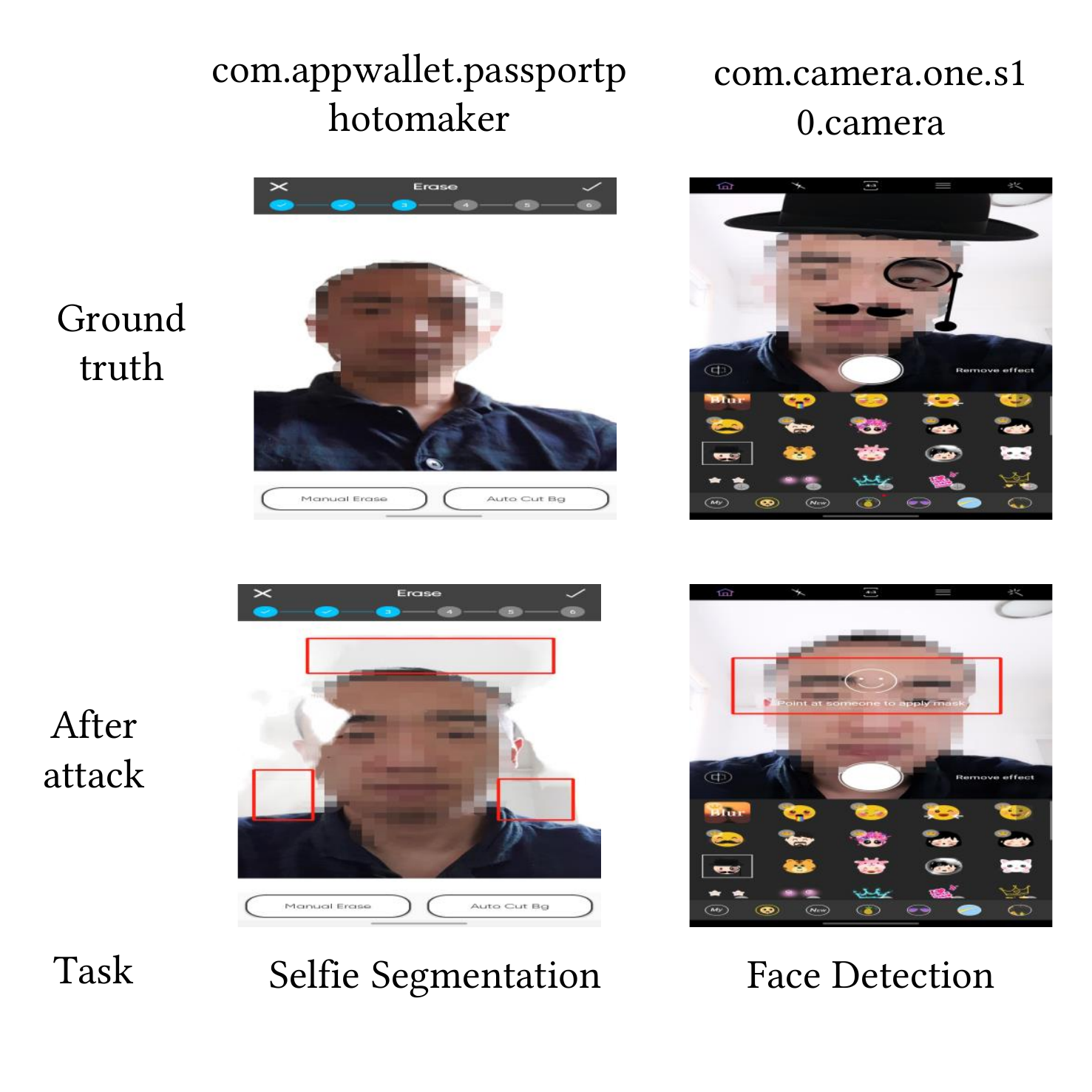}
   \caption{The behavior of two real-world DL apps in our rotation attack. \protect\footnotemark}
   \label{fig:8}
   \vspace{-0.2cm}
\end{figure}

\footnotetext{The person in this test is one of the authors}

\paragraph{Injecting Malicious Code to DL Apps and Evaluating the Attack Performance.}

By performing the pre-processing code location and malicious code injection against the identified DL apps, we find that 61.89\% (320/517) DL apps with TFLite models importing the MLKit package, and we successfully inject the malicious code into 81.56\% (261/320) DL apps using MLkit. Regarding the performance of our attack, we take two real-world DL apps with MLKit as examples to display the result, as depicted in Figure \ref{fig:8}. As observed, face detection and selfie segmentation cannot work when we modify the rotation values. In other words, its accuracy will be directly influenced by modifying the rotation values. 

%% comment by ye #2 3. % 0.74
% 0.29
% 0.03
% others such as pose recognization

\begin{figure}[htp]
  \centering
%   \fbox{\rule{0pt}{2in} \rule{0.9\linewidth}{0pt}}
   \includegraphics[width=0.95\linewidth]{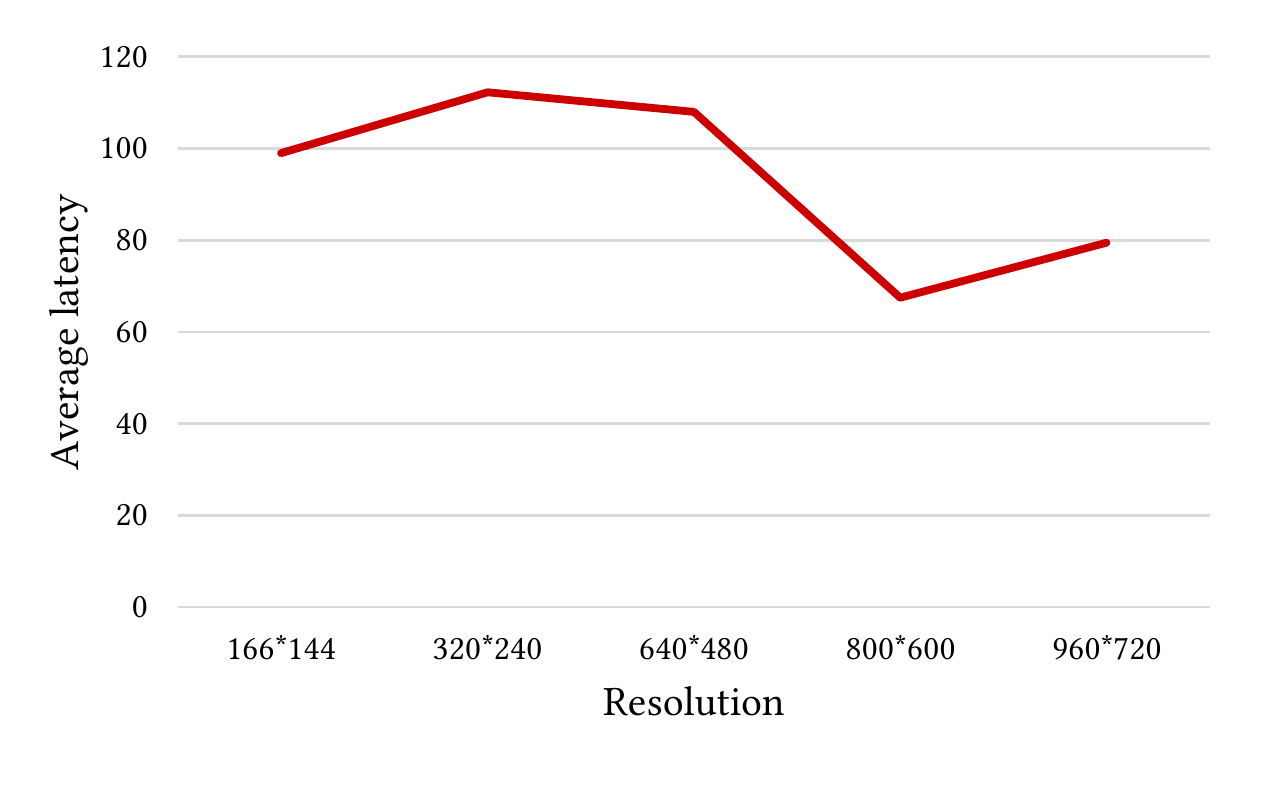}
   \vspace{-0.5cm}
   \caption{The impact of changing resolution on detection latency.}
   \label{fig:9}
\end{figure}

Regarding the research on latency, in order to avoid the influence brought by different logic in different DL apps, we directly use the sample code for face detection provided by MLKit, which contains the straightforward pre-processing, inference method invoking and post-processing codes. After that, the malicious codes are injected and installed on our testing Android phone to test the latency change in different situations, the result is shown in Figure~\ref{fig:9}. As seen, the change of height and width has impact on the detection latency of DL model.

\section{Discussion}
\paragraph{Result.} In this paper, we propose a new attack focused on data processing which aims to disturb the data pre-processing before the inference task in DL apps with TensorFlow DL models. Specifically, our attack cannot be influenced by the obfuscation, one of the most effective methods to protect apps, and could be applied into real-world Android apps. According to the experimental result, almost 61.89\% (320/517) DL apps with TF models adopted the MLKit, which could be directly used by developers to achieve AI services without training models. Furthermore, our attack successfully injects the malicious code into 81.56\% (261/320) DL apps. It could be observed that the modified resolution and rotation value could influence the inference latency and accuracy of DL models. The result also demonstrates that the larger the height and width are, the higher the detection latency will be. Meanwhile, our attack does not require the adversaries to know the inner structure of DL models, indicating that our attack is practical and dangerous.

\paragraph{Limitations.}
During the malicious code injection in Paragraph \pageref{malicious}, some apps cannot be installed after the injection. For example, if the developers encrypt an app through a complex algorithm, this kind of apps cannot be directly injected with the malicious code and re-compiled. The straightforward approach is to decrypt the original app and implement our approach. However, based on the time limitation, we only focus on those apps with simple encryption or without encryption. 

\paragraph{Future Work.}
In this work, we only focus on the DL apps using MLKit because it occupies the majority of DL apps with TensorFlow rather than those with customized TensorFlow DL models. It may be interesting to explore the pre-processing rules of these DL apps to make our attack more robust. Furthermore, the result generated from the inference method will also be processed to make it understandable by users, which process is called data post-processing. For example, the inference method of object detection will provide several probabilities of the target object. Then we could modify the probabilities weight to generate an incorrect prediction. In this case, we may further improve our attack by disrupting both data pre-processing and post-processing.

Furthermore, investigating the white-box scenario is imperative, as having knowledge of the inner structure of deep learning models or the obfuscation methods adopted by the developers could potentially offer greater flexibility in conducting attacks compared to the current research approach. For example, if we know the inner structure of the model, a more focused approach could be employed to perturb the preprocessing phase instead of solely focusing on resize or rotation adjustments. Similarly, upon knowing the obfuscated method of DL apps, the attack could initially de-obfuscate the code and subsequently inject malicious code based on the known obfuscations.

\section{Conclusion}
In this paper, we introduce a novel data pre-processing designed to target on-device deep learning models. Our approach is simple yet effective, as evidenced by experimental results that demonstrate its effectiveness against DL app across different categories. We also exemplify the potential impact of this attack through the real-world DL app. By showcasing the potential damage that this attack can cause, we aim to increase awareness within the research community regarding the need to protect on-device models. In general, our proposed attack serves as a valuable contribution to the growing body of research on securing deep learning models in mobile apps.

\begin{acks}
We express our gratitude to the anonymous reviewers for their valuable and constructive feedback. This work was supported in part by the National Key R\&D Program of China (No. 2021YFB2900100), and the National Natural Science Foundation of China (No. U22B2022,62002294).
\end{acks}

% \begin{acks}
% First, I would like to extend my sincere gratitude to my supervisor Dr. 
% Dr. Chunyang Chen and Dr. Yujin Huang, for their instructive advice and helpful suggestion. I am deeply grateful for 
% their help in completing my minor thesis. They spend much time 
% listening to my progress in each period and give their insightful 
% guidance.

% I also owe particular gratitude to all my roommates, from whom I 
% could always be encouraged and given practical advice when I 
% wanted to give up at each time.
\bibliographystyle{ACM-Reference-Format}
\bibliography{sample-sigplan}

% \appendix

% \section{CODES AND DATASETS}
% The DL app dataset and source code used in this paper are publicly accessible. We hide them for review purposes.

% \url{https://github.com/Alexender-Ye/attack-focused-on-pre-processing}

\end{document}